\documentclass[conference,a4paper]{IEEEtran}
\IEEEoverridecommandlockouts
\usepackage{cite}
\usepackage{amsmath,amssymb,amsfonts}
\usepackage{algorithmic}
\usepackage{graphicx}
\usepackage{textcomp}
\usepackage{xcolor}
\usepackage[nolist]{acronym}
\usepackage{paralist}
\usepackage{hyperref}
\def\BibTeX{{\rm B\kern-.05em{\sc i\kern-.025em b}\kern-.08em
    T\kern-.1667em\lower.7ex\hbox{E}\kern-.125emX}}

\begin{document}

\makeatletter
\def\ps@IEEEtitlepagestyle{%
\def\@oddfoot{\parbox{\textwidth}{\centering\normalsize{979-8-3503-9678-2/23/\$31.00 \copyright 2023 IEEE}\vspace{2em}}
}%
}

\makeatletter
\def\ps@IEEEtitlepagestyle{%
\def\@oddfoot{\parbox{\textwidth}{\footnotesize
Author's version of a paper accepted for publication in Proceedings of 2023 IEEE PES Innovative Smart Grid Technologies Europe (ISGT EUROPE). 
\\
\textcopyright{} 2023 IEEE. 
Personal use of this material is permitted.  
Permission from IEEE must be obtained for all other uses, in any current or future media, including reprinting/republishing this material for advertising or promotional purposes, creating new collective works, for resale or redistribution to servers or lists, or reuse of any copyrighted component of this work in other works.\vspace{1.2em}}
}%
}
\makeatother

\begin{acronym}
\acro{sg}[SG]{smart grid}
\acroplural{sg}[SGs]{smart grids}
\acro{der}[DER]{distributed energy resource}
\acroplural{der}[DERs]{distributed energy resources}
\acro{ict}[ICT]{information and communication technology}
\acro{fdi}[FDI]{false data injection}
\acro{scada}[SCADA]{Supervisory Control and Data Acquisition}
\acro{mtu}[MTU]{Master Terminal Unit}
\acroplural{mtu}[MTUs]{Master Terminal Units}
\acro{hmi}[HMI]{Human Machine Interface}
\acro{plc}[PLC]{Programmable Logic Controller}
\acro{dmz}[DMZ]{Demilitarized Zone}
\acroplural{plc}[PLCs]{Programmable Logic Controllers}
\acro{ied}[IED]{Intelligent Electronic Device}
\acroplural{ied}[IEDs]{Intelligent Electronic Devices}
\acro{rtu}[RTU]{Remote Terminal Unit}
\acroplural{rtu}[RTUs]{Remote Terminal Units}
\acro{iec104}[IEC-104]{IEC 60870-5-104}
\acro{apdu}[APDU]{Application Protocol Data Unit}
\acro{apci}[APCI]{Application Protocol Control Information}
\acro{asdu}[ASDU]{Application Service Data Unit}
\acro{io}[IO]{information object}
\acroplural{io}[IOs]{information objects}
\acro{cot}[COT]{cause of transmission}
\acro{mitm}[MITM]{Man-in-the-Middle}
\acro{fdi}[FDI]{False Data Injection}
\acro{ids}[IDS]{intrusion detection system}
\acroplural{ids}[IDSs]{intrusion detection systems}
\acro{siem}[SIEM]{Security Information and Event Management}
\acro{mv}[MV]{medium voltage}
\acro{lv}[LV]{low voltage}
\acro{cdss}[CDSS]{controllable distribution secondary substation}
\acro{bss}[BSS]{battery storage system}
\acroplural{bss}[BSSs]{battery storage systems}
\acro{pv}[PV]{photovoltaic inverter}
\acro{mp}[MP]{measuring point}
\acroplural{mp}[MPs]{measuring points}
\acro{dsc}[DSC]{Dummy SCADA Client}
\acro{fcli}[FCLI]{Fronius CL inverter}
\acro{fipi}[FIPI]{Fronius IG+ inverter}
\acro{sii}[SII]{Sunny Island inverter}
\acro{tls}[TLS]{Transport Layer Security}
\acro{actcon}[ActCon]{Activation Confirmation}
\acro{actterm}[ActTerm]{Activation Termination}
\acro{rtt}[RTT]{Round Trip Time}
\acro{c2}[C2]{Command and Control}
\acro{dst}[DST]{Dempster Shafer Theory}
\acro{ec}[EC]{Event Correlator}
\acro{sc}[SC]{Strategy Correlator}
\acro{ioc}[IoC]{Indicator of Compromise}
\acroplural{ioc}[IoCs]{Indicators of Compromise}
\acro{ot}[OT]{Operational Technology}
\acro{it}[IT]{Information Technology}
\acro{aucmorf}[AUC-MORF]{Area Under the Most Relevant First Perturbation Curve}
\acro{aupc}[AUPC]{Area Under Perturbation Curve}
\acro{ics}[ICS]{Industrial Control System}
\acroplural{ics}[ICSs]{Industrial Control Systems}
\acro{pera}[PERA]{Purdue Enterprise Reference Architecture}
\acro{dpi}[DPI]{Deep Package Inspection}
\end{acronym}

\bstctlcite{IEEEexample:BSTcontrol}

\title{Benchmark Evaluation of Anomaly-Based Intrusion Detection Systems in the Context of Smart Grids}

\author{
\IEEEauthorblockN{%
Ömer Sen\IEEEauthorrefmark{1}\IEEEauthorrefmark{2},
Simon Glomb\IEEEauthorrefmark{1},
Martin Henze\IEEEauthorrefmark{3}\IEEEauthorrefmark{4},
Andreas Ulbig\IEEEauthorrefmark{1}\IEEEauthorrefmark{2}
}

\IEEEauthorblockA{%
\IEEEauthorrefmark{1}\textit{IAEW, RWTH Aachen University,} Aachen, Germany |
\IEEEauthorrefmark{2}\textit{Digital Energy, Fraunhofer FIT,} Aachen, Germany\\
Email: \{o.sen, a.ulbig\}@iaew.rwth-aachen.de, simon.glomb@rwth-aachen.de, \{oemer.sen, andreas.ulbig\}@fit.fraunhofer.de}
\IEEEauthorblockA{%
\IEEEauthorrefmark{3}\textit{SPICe, RWTH Aachen University,} Aachen, Germany |
\IEEEauthorrefmark{4}\textit{CA\&D, Fraunhofer FKIE,} Wachtberg, Germany\\
Email: henze@cs.rwth-aachen.de, martin.henze@fkie.fraunhofer.de}
}

\IEEEoverridecommandlockouts

\maketitle

\IEEEpubidadjcol

\begin{abstract}
The increasing digitization of smart grids has made addressing cybersecurity issues crucial in order to secure the power supply.
Anomaly detection has emerged as a key technology for cybersecurity in smart grids, enabling the detection of unknown threats.
Many research efforts have proposed various machine-learning-based approaches for anomaly detection in grid operations.
However, there is a need for a reproducible and comprehensive evaluation environment to investigate and compare different approaches to anomaly detection.
The assessment process is highly dependent on the specific application and requires an evaluation that considers representative datasets from the use case as well as the specific characteristics of the use case.
In this work, we present an evaluation environment for anomaly detection methods in smart grids that facilitates reproducible and comprehensive evaluation of different anomaly detection methods.
\end{abstract}


\section{Introduction}
The evolution of power grids into \acp{sg} has brought changes to the \ac{ict} infrastructure.
Traditional \acp{sg} used to have isolated and proprietary components, creating barriers to external third-parties.
However, with the increasing penetration of \ac{ict}, the enhanced interconnection of \acp{sg} has resulted in new cybersecurity threats in the form of cyberattacks~\cite{pan2015developing}.
To better detect these threats, detection mechanisms are needed to observe and monitor the cybersecurity posture of \acp{sg}.
\acp{ids} have become a standard security component for identifying unusual activities and cyber-threats in the system~\cite{shun2008network}.
An \ac{ids} is a software designed to detect malicious activity on a network or system~\cite{wagle2021utilizing}.

Despite significant development over the years, existing \acp{ids} still face challenges in improving detection accuracy, reducing false positive rates, and detecting unknown attacks.
Machine learning has emerged as a crucial research topic to address these issues, as it has shown the potential to achieve high accuracy and generalizability in detecting new attacks~\cite{liu2019machine}.
However, these approaches often suffer from high false positive rates~\cite{samrin2017review} and lack explainability in issued alerts~\cite{kus2022false}.

To minimize the effort and expertise required for post-processing a large volume of issued alerts and reduce the time needed to correctly identify and understand the situation, an anomaly detection system with high precision and explainable alerts is necessary.
However, this requires an anomaly detection approach with reliable and explainable detection quality that needs to be verified and validated.

To evaluate the properties of such a system, experimental studies are necessary, where approaches are examined based on different criteria, scenarios (datasets), and use cases.
However, such studies often rely on the evaluation of a single dataset and lack the use of in-depth metrics to comprehensively evaluate a system~\cite{wolsing2022ipal}.
Thus, a reproducible and comprehensive benchmark environment is needed to support research in this direction.

In this work, we address the challenges of reproducibility of performance evaluation using different datasets, evaluation of the explanatory power of different machine learning approaches, and comparison of different anomaly detection systems in the same environment, with the goal of studying and evaluating machine-learning based anomaly detection for \acp{sg}.
Our contributions in this paper include presenting the current state of the art in machine learning testing and highlighting the benchmarking issues (Section~\ref{sec:analysis}), describing the overall benchmark approach covering dataset selection to performance evaluation of predictions (Section~\ref{sec:methodology}), and demonstrating and discussing the capabilities of our proposed approach through a case study (Section~\ref{sec:result}).

\section{Background \& Related Work}\label{background:ov}
In this section, we describe \ac{ids}, its function in the context of anomaly detection, and its main features in terms of explainability.
We also present related work in the context of machine learning-based \ac{ids} and benchmarking.

\subsection{Intrusion Detection Systems}\label{background:ids}
An \ac{ids} is a security solution, either software or hardware-based, that monitors and analyzes network or system activity to detect potential cyber threats like unauthorized access, misuse, or disruption.
It can be implemented in different ways, including individual devices or network-wide, and can monitor various data types like device logs, network traffic, and process data to identify suspicious or malicious activity.

Anomalies in \acp{ics} like \acp{sg} can be caused by various factors, including attacks on different segments, altered communication functions, and faults like protocol errors or missing messages.
Machine learning-based \acp{ids} use complex methods to detect anomalies, but explainability is an increasingly important feature to verify results and maintain confidence.
Explainability refers to the ability to provide qualitative understanding of the relationship between model input and prediction in a user-friendly manner%
Local explanations explain a model's behavior on one input instance, while global explanations understand the inner workings of a model's behavior.
Explainability methods can be model-agnostic or model-specific.

\subsection{Purdue Model}\label{background:purdue}
It is crucial to utilize realistic datasets that accurately reflect the complexity and diversity of real-world \ac{ics} environments when evaluating machine learning-based \acp{ids}.
The Purdue Reference Model, which is a component of \ac{pera}~\cite{williams1996overview}, provides a helpful framework for comprehending the various layers and components of \ac{ics} networks.
Hence, it can be utilized as a guide for selecting datasets for benchmarking purposes.

The Purdue Model is frequently utilized as a reference model for designing, analyzing, and securing \ac{ics} networks, as it helps understand the architecture and components of an \ac{ics}.
It splits the \ac{ics} architecture into two main zones: the \ac{ot} zone, which comprises the physical equipment and operational processes monitoring and control systems, and the \ac{it} zone, which encompasses the systems used for data management and communication.

The Purdue Model also includes a \ac{dmz} between the \ac{it} and \ac{ot} zones to control access between the two areas.
Therefore, the Purdue Model divides the \ac{ics} architecture into six levels, with each level representing a different layer of the \ac{ics} system.

\subsection{Related Work}\label{background:related_work}
The machine learning community has been exploring ways to create benchmark environments for evaluating the performance of different machine learning algorithms in detecting anomalies and intrusions in \ac{ics}.
One approach is to create benchmark libraries such as the Penn Machine Learning Benchmark~\cite{olson2017pmlb} and the Scientific Machine Learning Benchmark suite~\cite{thiyagalingam2022scientific}.
These libraries are intended to provide valuable resources for testing and comparing the performance of different algorithms.
Another approach is to create datasets from actual \ac{ics} environments to evaluate the performance of algorithms.
Researchers have proposed using datasets such as the Cyber-kit datasets~\cite{mubarak2021ics} and the Numenta anomaly benchmark~\cite{lavin2015evaluating} to evaluate unsupervised anomaly detection techniques or \acp{ids} in \ac{ics}.
In addition, studies have focused on the use of datasets specific to \acp{sg} for evaluating the performance of machine learning algorithms in detecting anomalies in \ac{sg} systems~\cite{bernieri2019evaluation}.

Several works have evaluated the accuracy and false positive rate of different anomaly detection approaches using datasets from real \ac{ics} environments~\cite{liyakkathali2020validating}.
Similarly, studies have proposed methods to assess the performance of machine learning algorithms in detecting anomalies in \acp{sg} and evaluate their performance based on various metrics such as accuracy, precision, and recall~\cite{mohammadpourfard2019benchmark}.
Overall, benchmarking and creating datasets are important approaches to evaluate the effectiveness of machine learning algorithms in anomaly detection for \ac{ics} and \acp{sg}.

\subsection{Problem Analysis}\label{background:problem_analysis}
The field of machine learning-based anomaly detection for \ac{ics} is rapidly growing and diverse, with many datasets and benchmarks proposed to evaluate machine learning algorithms for this purpose.
However, there is still a need for more specialized and comprehensive approaches that consider the unique characteristics and constraints of this domain.
Specifically, existing benchmarks may not cover all relevant evaluation methods or consider all important factors, such as the explainability of algorithms and the statistical confidence of findings. %
Furthermore, \acp{sg} may have unique characteristics in terms of their inputs, metrics, or other domain-specific factors, which are not always considered in benchmarks.
To address these issues, this paper aims to provide a comprehensive benchmark environment that evaluates machine learning-based anomaly detection for \acp{sg} using a diverse range of metrics and criteria, with a particular focus on the explainability of models.

\section{Analysis}\label{sec:analysis}
In this section, we analyze requirements of a benchmarking machine learning-based anomaly detection in \acp{sg}.

\subsection{Requirement}\label{sec:analysis_req}
The reproducibility of a research study refers to the ability to obtain the exact same results.
To ensure high reproducibility, various aspects need to be considered, such as datasets, data preprocessing, model training, model evaluation, and model software and hardware environment.
In addition to these requirements for individual studies, there are also requirements for benchmarks in general, including reporting, coverage, and extensibility.

The requirements include using publicly available datasets related to \ac{ics}, integrating common preprocessing methods, allowing customization of preprocessing methods and training data, comparing multiple algorithms, providing analysis for binary classification, using statistical methods, controlling randomization, providing a software environment as a container, and investigating new methods.
The environment should also provide information about the CPU and RAM, store necessary information for the reproduction of the study, and ensure that the datasets and software environment are accessible.

\subsection{Datasets}\label{sec:analysis_dataset}
By referring to the Purdue model (cf. Section~\ref{background:purdue}), datasets reflecting real-world attack scenarios can be used.
It also allows the selection of datasets that represent specific levels or components of the \ac{ics}, thus making it more relevant to evaluate and compare machine learning-based \acp{ids} for \ac{ics} environments.
These datasets with realistic and actual attacks enable accurate evaluation of \acp{ids}' performance and reflect real-world scenarios, making the benchmark more realistic and applicable for the \ac{ics} environment.

Various datasets are available, each useful for benchmarking different types of \acp{ids}.
The Bot IoT dataset includes anomalies related to botnets categorized as Level 0, which represent physical components of a system~\cite{koroniotis2020designing}.
The Power System dataset~\cite{MSU_ORNL_2014} focuses on anomalies at Level 1 and Level 2, which monitor and control physical equipment and operational processes, while the CICIDS-2017~\cite{sharafaldin2018toward} and CIDDS-002 datasets~\cite{ring2017creation} provide a range of intrusion and attack scenarios assigned to Levels 3 and 4.
The UNSW-NB15~\cite{moustafa2016evaluation} and NSL-KDD datasets~\cite{NSL-KDD_2023} are focused on anomalies at Levels 4 and 5, managing logistics of productive operations and providing communications and data
These datasets provide a diverse set of anomalies, enabling evaluation of \acp{ids} to detect and classify various types of network-based threats in \ac{ics}.

To evaluate machine learning-based \acp{ids} in \ac{ics}, including \ac{sg} research, it is important to have a benchmark environment that includes a diverse set of anomalies representative of the type of attacks that these systems may encounter in real-world scenarios.
Selecting datasets that include anomalies representative of the different layers of the Purdue Model ensures that the machine learning-based \acp{ids} are tested against a diverse set of attack scenarios that are relevant to \acp{ics}.
Publicly available, labeled, and network-based datasets are considered for the evaluation of multiple datasets.
Additionally, it is critical to have explainable methods to understand the reasoning behind an \ac{ids} decision, building trust in these systems, and mitigating the impacts of false positives.

\subsection{Machine Learning Model}\label{sec:analysis_ml}
The paper by Davis et al.~\cite{davis2011data} lists several important preprocessing steps for network-based \acp{ids}.
These steps include data transformations, data cleaning, data reduction, discretization, and re-labeling.
However, feature construction and dataset creation are excluded as they are highly dependent on the specific protocols and features of the datasets, and this work aims to abstract from such specifics.
It is recommended to consider all other preprocessing steps, and multiple preprocessing methods should be usable without impairing other methods, i.e. allowed to be omitted in evaluation.

Ensemble learning, which combines multiple learning algorithms, is often used in recent works to improve classification performance.
In this paper, the term ''stacking'' is used synonymously with ensemble methods, and the models being combined are referred to as first-level models, while the model that combines their predictions is the second-level model.
For the proposed framework, any stacking algorithm should be analyzable, and it should be possible to choose the number and type of base classifiers arbitrarily.
Additionally, we want to be able to compare two arbitrary algorithms, and the data on which the base classifiers are trained should be customizable.

Evaluation metrics for \acp{ids} are highly dependent on the specific application.
High detection accuracy, runtime efficiency, and explainability are important criteria, and enhancing the robustness of \acp{ids} is a novel way to improve detection accuracy.%
Furthermore, diversity in ensemble methods is a fundamental issue and should be evaluated. %
The proposed framework should support research on these main directions for \acp{ids}.
Binary classification is chosen as the evaluation task, and multiple datasets should be used for evaluation.
Statistical tests and confidence intervals should be used to quantify the generalizability of results in a reproducible manner (i.e. control of randomization). %

\subsection{Environment}\label{sec:analysis_env}
To ensure reproducibility, it is recommended to use containers, as stated in~\cite{tatman2018practical}.
We require that the container used in the proposed environment should be versioned to make the environment accessible and enable the creation of reproducible results.
It is also important to report hardware information, specifically CPU and RAM, as relevant hardware information. %
This is necessary to assess studies conducted on different devices and potentially use them in the future.

All information pertaining to each study must be stored in a database to ensure reproducibility.
This includes information about the dataset, preprocessing strategy, model training with hyperparameters, training method, model evaluation, as well as software and hardware environment.
Our goal is to ensure that the reporting of each examination is reproducible and provides the context for reproducing the results.

Our work is limited to anomaly-based network \acp{ids} with datasets specific to the \acp{sg} domain.
We focus exclusively on binary classification and machine learning approaches, both supervised and unsupervised.
However, we require the investigation of new methods, and the implementation and evaluation of new approaches should be possible.
We will present a methodology as a reference implementation, which incorporates the requirements we have analyzed.
\section{Methodology}\label{sec:methodology}
In this section, we present a concrete methodology on how the requirements can be fulfilled.
With the methodology, we also present how our tool is constructed and what it provides.

\subsection{Overview}\label{sec:methodology_overview}
\begin{figure}
    \centering
    \includegraphics*[scale=1, width=0.5\textwidth]{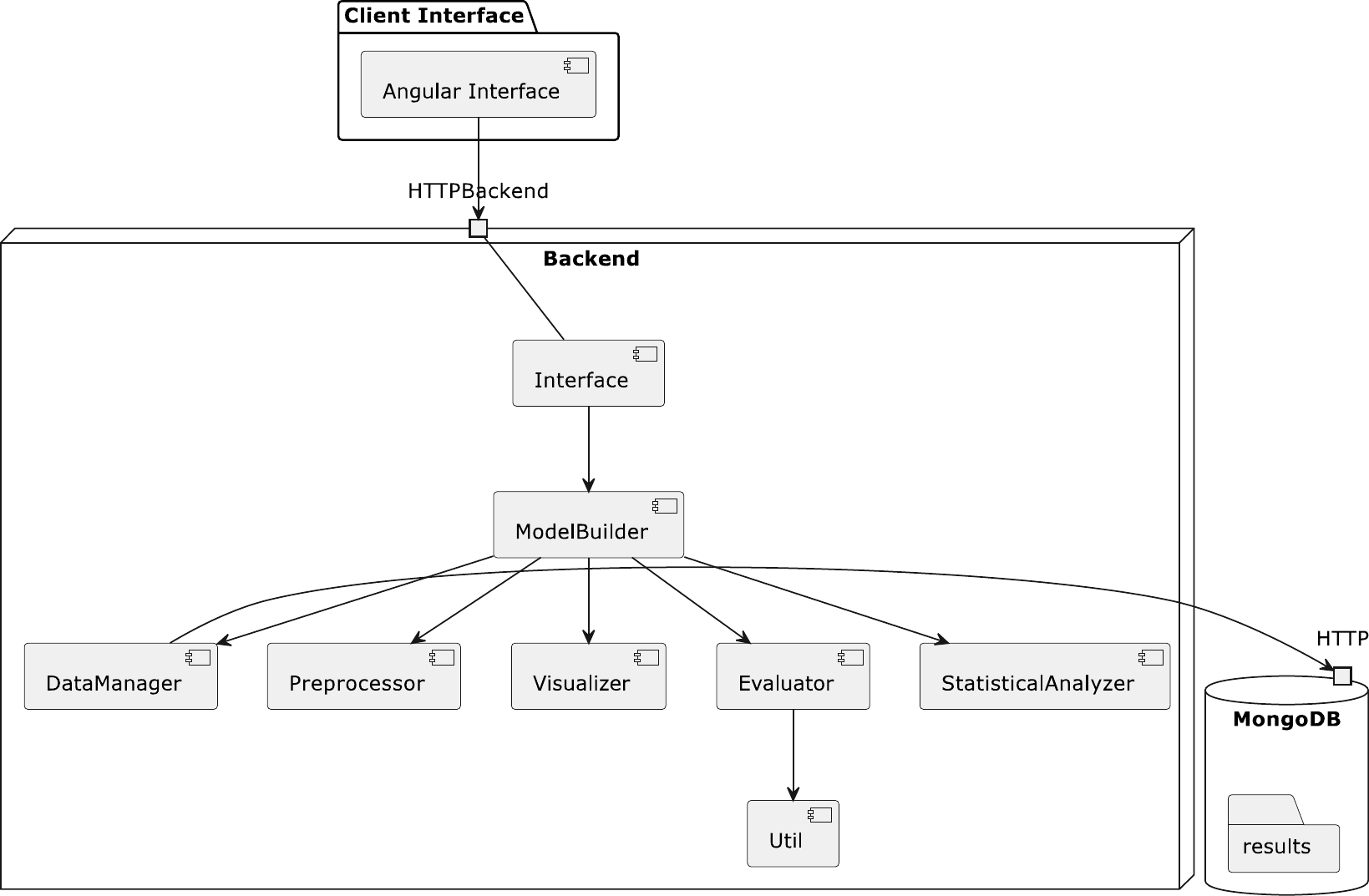}
    \caption{UML Component-Diagram for the Core Components.}
    \label{component_diag}
\end{figure}

We present a component diagram in Figure~\ref{component_diag} which provides an overview of our design.
Each methodology component can be associated with a specific component, thereby offering a tool that meets our established requirements for evaluating \acp{ids}.
Because machine learning algorithms can have hyper-parameters that are arbitrarily nested, a relational database is unsuitable for this purpose.
Therefore, we have chosen to use a document-oriented database to store the results from the environment, specifically the MongoDB database management system.
BSON (Binary JSON) is the format for data storage, which is a superset of JSON that includes additional data types.
Additionally, we provide a BSON-Schema to establish the structure for the results.
The schema includes JSON-LD to provide linked data for the results, which we have used as an example and as a proof of concept that the proposed architecture is fully compatible with JSON-LD.
\subsection{Preprocessing and Evaluation Strategy}\label{sec:methodology_pre_ev}
In the following section, we will explain how we preprocess our datasets prior to training, and how we select the appropriate approach for assessing the explainability of our method.

The first step in preprocessing is data transformation.
Normalization is considered to be a crucial technique to enhance model performance and computational efficiency for network-based \acp{ids}. %
Moreover, we need to perform discretization, as symbolic features must be transformed into numerical features.
As we cannot guarantee that the features are ordinal, we will use the commonly used One-Hot-Encoding method.
For data reduction, we will focus on dimensionality reduction, and we will choose the widely used principal component analysis (PCA) method. %

We have chosen the SHAP framework to generate explanations~\cite{van2022tractability}.
SHAP can explain the output of any machine learning model, and is based on game-theoretic concepts.
The reason for selecting SHAP is its ability to produce global explanations, which can help to build trust in the \ac{ids}.
Additionally, we deemed it necessary for the method to also generate local explanations for specific intrusion detection cases, as this can help to improve trust in the \ac{ids} when used in practice. %
This is because the same method can be used to evaluate a model and a particular case when the system is put into practice.
Another mandatory requirement was that the method had to be model-agnostic, as this allows us to apply the method to all models and are not restricted on the \acp{ids} that can be analyzed. %
Furthermore, SHAP is already being studied for use in \acp{ids}. %

The metrics typically used to measure the performance of \acp{ids} include training time, prediction time, and explanation time.
The literature suggests using the following metrics for classification in the intrusion detection context: False Positive Rate, AUC Score, Balanced Accuracy, and MMC.
In the following, we will cover additional metrics for evaluating anomaly detection.

\subsection{Measuring Explainability}\label{sec:methodology_exp}
We will evaluate the quality of explanations based on known properties as described in~\cite{islam2020towards}.
Our evaluation will cover three aspects: accuracy, stability, and degree of importance.
Importantly, these quality measures are not specific to any particular explanation method, including SHAP.
For SHAP, we will compute the SHAP values using test data as suggested by~\cite{molnar2020interpretable}, which is relevant for computing all the explanation metrics.

To measure the accuracy of explanations, we aim to quantify the discrepancy between the output of a classifier predicted by explanations and the actual output of the model.

For measuring the stability of explanations, we will compute the maximum sensitivity of the explanation to input perturbations as suggested in~\cite{vanni2022design}.
The sensitivity should be as low as possible, and we will use the definition and approach provided by~\cite{kakogeorgiou2021evaluating}, which can be found in Equation~\ref{eq:sens_max}.
\begin{equation}
    \operatorname{SENS}_{\text{MAX}}(\Phi, f_{c}, x, r) =
    \max_{\substack{\|x^{\prime}-x\|_{\infty} \\ \leq r}}
    \|\Phi(f_{c}, x^{\prime}) - \Phi(f_{c}, x)\|_{F}
    \label{eq:sens_max}
\end{equation}
Here, $|\cdot |F$ denotes the Frobenius norm, $f_{c}$ is the classifier, $\Phi$ is the explanation method, $x$ is the unlabeled data, and $r$ is the perturbation radius.
We found a perturbation radius of 0.01 suitable for normalized data through experiments.
\begin{equation}
    AUC_{\text{MoRF}}(\Phi, f_{c}, x, r) =
    \sum_{k=2}^{D} \frac{f_{c}(x_{\text{MoRF}}^{(k-1)}) + f_{c}(x_{\text{MoRF}}^{(k)})}{2}
    \label{aupc_3}
\end{equation}
We will also use the \ac{aucmorf} metric to evaluate the degree of importance of the features declared by the explanation $\phi(f_c, x)$.
The metric measures how fast the output of the classifier decreases as we progressively remove information from the input x, based on the importance declared by the explanation.
We will use the definition of \ac{aupc} from~\cite{kakogeorgiou2021evaluating}.
The \ac{aucmorf} metric is computed by perturbing the most important features of the data $x^{(k)}_{MoRF}$, where $k$ denotes the number of most important features perturbed.
We will use the formula presented in Equation~\ref{aupc_3} to compute the area under the perturbation curve.
The smaller the area under the curve, the more reliable the information is, as it confirms that the declared most important features are indeed the most important ones.

\subsection{Measuring Robustness}\label{metric:robustness}
The stability of explanations has been measured using the concept of perturbations and sensitivity. %
Both classifiers' decisions and interpretability methods are sensitive to input perturbations and susceptible to adversarial attacks. %
Adversarial robustness, which represents the worst-case scenario for any unforeseen randomness, is focused on instead of robustness to random noise. %
The robustness we investigate is the robustness of an already learned classifier against evasion attacks. %
Poisoning attempts, where malicious training data is injected into the learning system, are not included in our measurement. %
The adversarial robustness of a data instance $x$ on the classifier $f_c$ is defined as the minimum perturbation $r$ that causes $f_c$ to misclassify $x$. %
\begin{equation}
    \Delta_{\mathrm{adv}}(x ; f_{c}) = \min_{\substack{r \in \mathbb{R}^{d} \\ f_{c}(x) - f_{c}(x+r) \neq 0}} \|r\|_{2}
\end{equation}
We measure the sensitivity of the classifier using the Lipschitz constant, which is the maximum ratio between output and input variations of $f_c$. %
\begin{equation}
    L(f_{c}) = \max_{\substack{x_{i}, x_{j} \\ \in \mathcal{T}}} \frac{\|f_{c}(x_{i})-f_{c}(x_{j})\|_{2}}{\|x_{i}-x_{j}\|_{2}}
    \label{eq:adversarial_lipschitz}
\end{equation}
We use a lower bound for the true global Lipschitz constant, and it is analyzed with respect to the $L_2$-norm. %
Other measures and approximations of the Lipschitz constant are available, but they are model-specific and, therefore, disregarded. %

\subsection{Statistical Methods}\label{meth:statistical_methods}
This section focuses on the utilization of statistical methods for quantifying the generalizability of results.
The work addresses the scenario of statistical comparison across multiple datasets, where sample size refers to the number of datasets used. %
Nonparametric tests are preferred in this case due to the lack of assumptions about homogeneity of variance. %
When comparing learning algorithms, the Wilcoxon signed ranks test is recommended for two-algorithm comparisons, while the Friedman test is suggested for comparisons involving more than two classifiers. %
Since our evaluation requires comparing two arbitrary algorithms, we will use the Wilcoxon signed ranks test.
However, it cannot be computed if all data points for both groups are equal.

Estimation statistics, which uses effect sizes and confidence intervals, can be used in addition to traditional hypothesis testing and is sometimes recommended to replace null hypothesis significance testing. %
Effect size is used to quantify the magnitude of an effect in our case of comparing algorithm metrics, and various measurements such as percentage, median, correlation, or Cohen's d (recommended for comparing learning algorithms) can be used as effect size.%
A value of 0.2 or lower is considered a small effect, 0.5 is considered medium, and 0.8 or higher is considered large for Cohen's d.
We will use this terminology in our scenario evaluation.
Cohen's d can have values in the range of $(-\infty,\infty)$, but it cannot be computed if both classifiers have no deviation. %

Confidence intervals are used to estimate the range of an unknown parameter.
In our case, we are interested in confidence intervals for the difference of sample means.
A sample size of 30 is often suggested as the minimum requirement for assuming a normal distribution of mean measurements. %
However, since this assumption does not hold for our analysis, we will use bootstrap confidence intervals~\cite{efron1994introduction} based on the implementation described in~\cite{ho2019moving}.

The Gardner-Altman estimation plot~\cite{gardner1986confidence} is considered suitable and is a standard plot for utilization with estimation statistics.
It presents all data points as a swarm plot to display all observed values, and shows the point estimate (e.g. mean difference) with its 95\% bootstrap confidence interval on a separate difference-axis.
Figure~\ref{fig:estimation_plot} shows a Gardner-Altman estimation plot for the mean difference of explanation sensitivity in a comparison of two explanation methods applied on ensembles.

\section{Result}\label{sec:result}
In this section, we present the general procedure of using our environment, and then present a case study on the investigation of a method for the use in the intrusion detection.

\subsection{Objective}\label{sec:result_case_study}
We have chosen the aspect of explainability from the topics we have covered, and demonstrated that our evaluation goes beyond classical metrics.
Moreover, we have shown that our evaluation can be more comprehensive than evaluating on a single dataset.
When investigating a specific \ac{ids}, it is useful to compare it with another system.
We abstract from the entire \ac{ids} and only consider the underlying detection method.
The user inputs two learning algorithms that must inherit the interface from the sklearn class \textit{sklearn.base.BaseEstimator}.
Without using our environment, there would be more work to do to evaluate the algorithms.
We encapsulate the knowledge presented in the methodology to facilitate comparable research results.
The environment adds value by providing a state-of-the-art evaluation process that reduces the workload for researchers.
New insights can be incorporated into this environment by extending the presented components.

\subsection{Investigation Subject}\label{sec:result_subject}
We investigate a new method called basic join, which focuses on the explainability of ensembles of classifiers.
We aim to explore the possibility of utilizing this method for intrusion detection systems in \acp{sg} using the proposed environment.
The issues that our proposed environment addresses are as follows: dataset integration, which is usually time-consuming but made easier with the proposed environment;
selection of domain-specific metrics and evaluation measures, which are already done and do not need to be re-implemented;
storage of results in a reproducible way for future re-evaluation without loss of configurations;
and provision of confidence estimation of results for the domain of \ac{sg}.
The basic join, which we want to evaluate, provides an explanation for the entire model.
Unlike the traditional approach of treating the whole model as a black box, the basic join determines the importance of features of all first-level models and then computes the dot product with the vector containing the importance of features of the second-level model.
The features of the second-level model are the predictions of the first-level models, so the importance can be seen as the contribution of the first-level models to the second-level output. %
As models, we chose two ensembles, one using the traditional approach and the other using the basic join.
Both ensembles consist of three learning algorithms: Logistic Regression, Decision Tree, and Multi-Layer Perceptron.
For preprocessing, we used normalization and did not include methods like principal component analysis as it impacts explainability. %
Our models will be evaluated on all included datasets using the metrics proposed in the methodology, with a focus on the explainability aspect.

\subsection{Case Study}\label{sec:result_findings}
\begin{figure}[h]
    \centering
    \includegraphics[scale=1, width=0.5\textwidth]{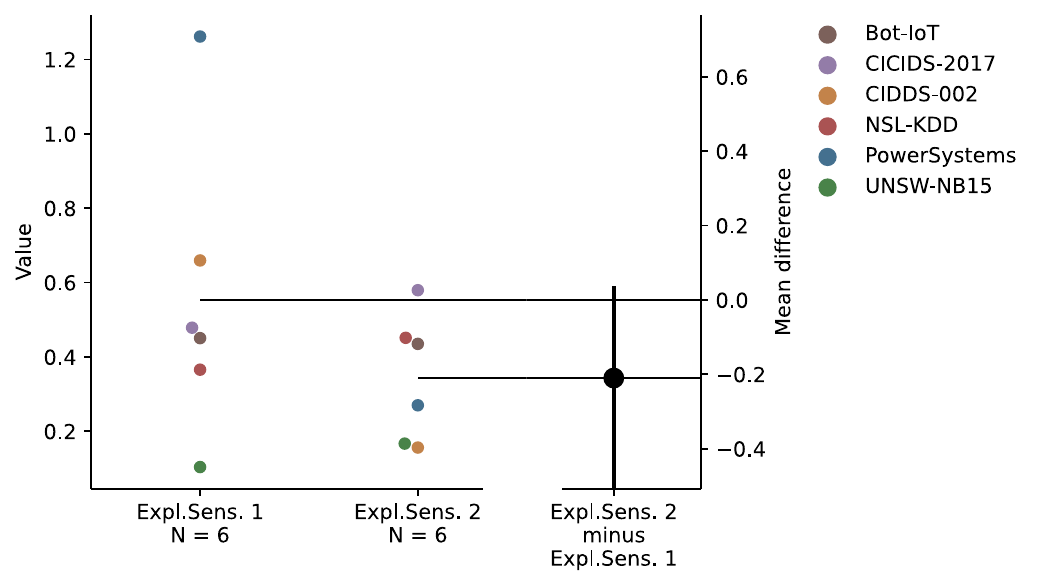}
    \caption{Gardner-Altman Estimation Plot for the Mean Difference of the Explanation Sensitivity (Expl.Sens.) for two Ensembles (both consisting of the same Logistic Regression, Decision Tree and Multi-Layer Perceptron).}
    \label{fig:estimation_plot}
\end{figure}
Regarding the explainability metrics, we observed the following: There was a small difference in explanation error (Cohen's d = -0.14).
No difference was observed in the area under the precision-recall curve (\ac{aupc}), which means that at least the first five features have the same order of relevance for both approaches (as we chose to evaluate five features for \ac{aupc}).
The most significant effects were seen in explanation time (Cohen's d = -0.35) and explanation sensitivity (Cohen's d = 0.31).
We present a Gardner-Altman Estimation plot for the latter metric in Figure~\ref{fig:estimation_plot}.
From the results, we can see that on average, the traditional approach has slightly lower explanation error and requires less time for explanation creation compared to the basic join.
However, the new method appears to have benefits in terms of the robustness of explanations on some datasets and may be advantageous for dealing with data noise.
The datasets where the basic join performs better are the Power System dataset and the CIDDS-002 dataset.
By utilizing our proposed environment, we can infer that the method under scrutiny has a lower susceptibility to noise, but it has certain downsides concerning computational resources.
Overall, the investigated method has this trade-off between noise sensitivity and computational requirements.

\section{Conclusion} \label{sec:conclusion}
\ac{sg} cybersecurity is crucial for securing \ac{sg} tasks, and anomaly detection is a critical technology in addressing this challenge.
This study introduces an evaluation environment for anomaly detection methods in \ac{sg} that allows for reproducible and comprehensive evaluation of different techniques, providing a general framework for evaluations and comparisons.
Our contributions to the energy community include conducting a detailed analysis of benchmark environment requirements for anomaly-based network \ac{ids}, providing a reference implementation for building such an environment for the energy sector by aggregating results from various research branches, and demonstrating a proof of concept for using statistical methods for cross-dataset evaluation in the context of \ac{ids}.
Future work includes incorporating measures of classifier diversity and combining datasets to enhance evaluation metrics, towards standardized benchmark environments for evaluating \acp{ids}.


\end{document}